\documentclass[12pt]{article}
\usepackage{color}
\usepackage{amssymb,amsmath,comment}
\usepackage{graphicx}
\usepackage{braket}
\usepackage[dvipdfmx]{hyperref}
\usepackage{epsfig}
\usepackage{here}
\graphicspath{{./Figures/}}
\newcommand{\ba}{\begin{align}}
\newcommand{\ea}{\end{align}}

\def\bea{\begin{eqnarray}}
\def\eea{\end{eqnarray}}
 \makeatletter
\def\alt{\mathrel{\mathpalette\gl@align<}}
\def\agt{\mathrel{\mathpalette\gl@align>}}
\def\gl@align#1#2{\lower.6ex\vbox{\baselineskip\z@skip\lineskip\z@
\ialign{$\m@th#1\hfil##\hfil$\crcr#2\crcr\sim\crcr}}} \makeatother
\renewcommand{\thefootnote}{\fnsymbol{footnote}}
\newcommand{\doi}[1]{\href{https://doi.org/#1}{doi:#1}}
\begin{document}
\begin{flushright}
\end{flushright}
\vspace*{1.0cm}

\begin{center}
\baselineskip 20pt 
{\Large\bf 
Detectable dimension-6 proton decay \\
in SUSY SO(10) GUT
at Hyper-Kamiokande
}
\vspace{1cm}

{\large 
Naoyuki Haba,
\ Yukihiro Mimura
\ and \ Toshifumi Yamada
} \vspace{.5cm}

{\baselineskip 20pt \it
Institute of Science and Engineering, Shimane University, Matsue 690-8504, Japan
}

\vspace{.5cm}

\vspace{1.5cm} {\bf Abstract} \end{center}

In the minimal SUSY SU(5) GUT
 with $O(1)$~TeV SUSY particles and $O(1)$ or below self-coupling for the GUT-breaking Higgs field,
 the width of the dimension-6 proton decay is suppressed below the reach of Hyper-Kamiokande.
In this paper, we point out that a SUSY SO(10) GUT which adopts only
 ${\bf 45}_{\rm H}+{\bf 16}_{\rm H}+{\bf \overline{16}}_{\rm H}$ GUT-breaking Higgs fields
 leads to an enhanced dimension-6 proton decay width detectable at Hyper-Kamiokande.
The enhancement is because the SU(5)-breaking VEV of ${\bf 45}_{\rm H}$ arises due to Planck-suppressed terms,
 $W\propto ({\bf 45}_{\rm H}^2)^2/M_*+{\bf 45}_{\rm H}^4/M_*$, and is therefore substantially larger than the other VEVs that conserve SU(5).
As a result, the $({\bf 3},{\bf 2},1/6)$ GUT gauge boson mass is about $1/5$ smaller than the $({\bf 3},{\bf 2},-5/6)$ GUT gauge boson mass
 and can induce a fast dimension-6 proton decay.
Through a numerical analysis on threshold corrections of the GUT gauge bosons and the physical components of the GUT-breaking Higgs fields, we confirm that the dimension-6 proton decay can be within the reach of Hyper-Kamiokande.

\thispagestyle{empty}

\newpage
\renewcommand{\thefootnote}{\arabic{footnote}}
\setcounter{footnote}{0}
\baselineskip 18pt
\section{Introduction}

The dimension-6 proton decay is an important prediction of the grand unified theory (GUT)~\cite{Langacker:1980js}.
The Super-Kamiokande experiment currently gives the 
bound of the partial proton lifetime
$\tau (p\to \pi^0 e^+) > 1.6 \times 10^{34}$~years (90\% confidence level)~\cite{Miura:2016krn},
and it will be searched
up to $6.3 \times 10^{34}$~years at $3\sigma$ level
by a 10~year exposure of one 187 kton fiducial volume detector
at Hyper-Kamiokande (HK)~\cite{Abe:2018uyc}.
Now that the HK experiment is scheduled to start in 2026,
 it is time to survey GUT models which predict the dimension-6 proton decay within the discovery reach.

In the minimal supersymmetric (SUSY) SU(5) GUT  \cite{Dimopoulos:1981zb,Dimopoulos:1981yj,Sakai:1981gr}
with $O(1)$~TeV SUSY particles,
 the partial lifetime for the dimension-6 proton decay via GUT gauge boson exchange is predicted to be more than a few times $10^{35}$~years naively.
The gauge coupling unification condition does not directly give the mass of the GUT gauge boson, because the mass of the physical components of 
 the SU(5)-breaking Higgs field {\bf 24}$_{\rm H}$ cannot be determined theoretically.
The GUT gauge boson mass becomes heavier (the proton lifetime is longer) if the self-coupling of {\bf 24}$_{\rm H}$ is smaller.
\footnote{
In non-SUSY SU(5) GUTs, some choices of GUT Higgses yield models that survive the SK bound and will be explored at HK \cite{Haba:2018vvu}.
}
The SUSY GUTs also predict the dimension-5 proton decay via colored Higgs exchange~\cite{Weinberg:1981wj},
 such as $p\to K^+ \bar \nu$,
whose current bound reads $\tau (p \to K^+ \bar\nu) > 5.9 \times 10^{33}$~years~\cite{Abe:2014mwa}
and which often gives a severe constraint on the model construction.
There are several ways to suppress the dimension-5 decay to a harmless level,
 e.g., by enhancing the colored Higgs mass with SUSY particle threshold with large wino/gluino mass ratio~\cite{Tobe:2003yj}, 
 with non-renormalizable superpotential of adjoint representations \cite{Bajc:2002pg}, or
with GUT particle thresholds in non-minimal models for the gauge coupling evolutions \cite{Dutta:2007ai}.
Other ways include assuming heavy squarks, or utilizing a cancellation among multiple Higgs couplings.
Compared to the dimension-5 decay, the dimension-6 proton decay involves less parameters and its naive prediction is above the current experimental bound.
Therefore, it is worth pursuing the possibility that $p\to \pi^0 e^+$ will be observed at HK.
In fact, as the LHC results imply that the SUSY particles have mass above multi-TeV scale,
 some people revisit the unification conditions in the context of 
the high-scale SUSY scenario~\cite{Hisano:2013cqa,Bajc:2016qcc,Ellis:2016tjc,Pokorski:2017ueo}.
As the wino and gluino are heavier, the unification scale becomes lower,
and it can reach the discovery range of HK for $\sim$10-100~TeV wino and gluino masses.

%
%

What about the dimension-6 proton decay in SUSY SO(10) GUTs?
The breaking pattern of the SO(10) symmetry has room for the existence of intermediate scales,
 and thus the prediction of the dimension-6 proton decay varies in a wide range.
Among various choices of the Higgs representations to break SO(10) to the SM gauge symmetry,
the simplest choice is ${\bf45}_{\rm H}+{\bf 16}_{\rm H}+ \overline{\bf16}_{\rm H}$,
 which is also the most economical in view of the total contribution to the beta coefficient for gauge couplings. 
The above choice of the Higgs representations gives characteristic vacua where 
 the GUT gauge boson with SM charge $({\bf 3},{\bf 2},1/6)$, which is absent in SU(5) GUT,
 is about $1/5$ lighter
 than the GUT gauge boson with SM charge $({\bf 3},{\bf 2},-5/6)$, which is also present in SU(5) GUT.
In the vacua, therefore, the dimension-6 proton decay width is enlarged compared to the minimal SU(5) model
 due to the exchange of the light $({\bf 3},{\bf 2},1/6)$ gauge boson.
So, it is worth scrutinizing the prediction of the dimension-6 proton decay in the above model, since the predicted proton lifetime can be in the range of HK.
To our best knowledge,
 this simple SO(10) model
 has not been investigated in light of 
 experimental accessibility of the dimension-6 proton decay.
In this paper, we will show a numerical calculation of the dimension-6 proton decay $p\to\pi^0e^+$ in the SO(10) model with
 ${\bf45}_{\rm H}+{\bf 16}_{\rm H}+ \overline{\bf16}_{\rm H}$ GUT-breaking Higgs fields.

We also find that in the characteristic vacua of the above model,
 the colored Higgs mass is enhanced by about 576 compared to the minimal SU(5) model
 due to threshold corrections of GUT gauge bosons and physical components of GUT-breaking Higgs fields.
\footnote{
Considering this enhancement and uncertainty of the Yukawa coupling unification, we omit an analysis of the dimension-5 decay in this paper.
}
So, this SO(10) model exhibits an interesting tendency that the dimension-6 decay width is enhanced and the dimension-5 decay width is suppressed.

This paper is organized as follows:
In Section 2, we present the spectrum of the SO(10) gauge bosons that gain mass
via symmetry breaking with ${\bf 45}_{\rm H}+{\bf 16}_{\rm H}+ \overline{\bf16}_{\rm H}$,
and show that the dimension-6 proton decay width can be enlarged.
In Section 3, we study how the characteristic SO(10) breaking vacua of the model are obtained.
In Section 4, GUT-scale threshold corrections are evaluated for the calculation of the dimension-6 proton decay width.
In Section 5, a detailed numerical result for the proton lifetime is presented.
Section 6 is for the conclusion.
In Appendix A, we show the mass spectrum 
of the multiplets which come from ${\bf 16}_{\rm H}+ \overline{\bf16}_{\rm H}$.
In Appendix B, an alternative, renormalizable superpotential with ${\bf 54}_{\rm H}$ 
representation GUT Higgs field is shown.







\section{Spectrum of the SO(10) gauge bosons and proton decay }

There are many ways to break the SO(10) gauge symmetry to
the SM gauge symmetry.
The most economical choice of the Higgs representations 
to break SO(10) is
${\bf 45}_{\rm H} + {\bf 16}_{\rm H}$.
We also introduce $\overline{\bf 16}_{\rm H}$ for $D$-flatness.
${\bf 45}_{\rm H}$ contains two SM singlets:
a SU(5) singlet ($a_1$) and a SM singlet in SU(5) adjoint ($a_{24}$).
By general vacuum expectation values (VEVs) of $a_1$ and $a_{24}$
(without particular relations between them),
SO(10) is broken down to $SU(3)_c \times SU(2)_L \times U(1)_Y \times U(1)^\prime$.
The SM singlet in ${\bf16}_{\rm H}$ ($v_R$), which is a SU(5) singlet, develops a VEV to break $U(1)^\prime$ 
 (the singlet in $\overline{{\bf16}}_{\rm H}$ ($\bar{v}_R$) also gains a VEV with $|v_R| = |\bar v_R|$ to keep D-flatness).
Due to the absence of cubic term of ${\bf 45}_{\rm H}$,
$a_{24}$ cannot acquire a VEV
 from the renormalizable superpotential.
However, by introducing non-renormalizable, quartic terms of ${\bf45}_{\rm H}$,
$a_{24}$ can acquire a VEV to break SU(5).

The GUT gauge boson masses generated by the VEVs of 
${\bf45}_{\rm H}$, ${\bf16}_{\rm H}$ and $\overline{\bf16}_{\rm H}$ are
\begin{eqnarray}
M^2_{X}/g_5^2 &=& \frac56 |a_{24}|^2 , \label{mx}\\
M^2_{Q}/g_5^2 &=& \frac45 \left|a_1 + \frac1{2\sqrt6} a_{24}\right|^2 +|v_R|^2,\\
M^2_{U}/g_5^2 &=& \frac45 \left|a_1 -\sqrt{\frac23} a_{24}\right|^2 + |v_R|^2,\\
M^2_{E}/g_5^2 &=& \frac45 \left|a_1 + \sqrt{\frac32} a_{24}\right|^2 + |v_R|^2,
\end{eqnarray}
where $X$, $Q$, $U$ and $E$
denote SO(10) gauge bosons whose SM charges are
$X: ({\bf3},{\bf2},-5/6)$, $Q: ({\bf3},{\bf2},1/6)$, $U: ({\bf3},{\bf1},2/3)$ and $E:({\bf1},{\bf1},1)$,
and $a_1$, $a_{24}$ and $v_R$ are the VEVs of canonically-normalized SM singlets.
The extra U(1) gauge boson mass is $g_5 \sqrt{5/2} v_R$.

When $a_{24} \gg a_1, v_R$,
we obtain
\begin{equation}
M_X : M_Q : M_U : M_E \simeq 5 : 1: 4 : 6.
\label{gaugebosonmasses}
\end{equation}
This ratio is easily obtained by the rule $(3+2): (3-2): (2+2): (3+3)$.

The dimension-6 proton decay operators are generated
not only by the $X$ gauge boson exchange but also by the
$Q$ gauge boson exchange.
The partial width of the dimension-6 proton decay is given by
\begin{equation}
\Gamma \propto A_{ R}^2 \left(\frac{1}{M_X^2}+\frac1{M_Q^2}\right)^2 + \frac{4A_{ L}^2}{M_X^4},
\end{equation}
where $A_{ L,R}$ are the renormalization factors for
$q \ell (u^c)^\dagger (d^c)^\dagger$ and $q q (e^c)^\dagger (u^c)^\dagger$ operators.
One finds that
the $Q$ gauge boson exchange gives much larger contribution 
 when $M_X:M_Q \simeq 5:1$.
The ratio of the decay width in SU(5) GUT ($M_Q \to \infty$)
and in the SO(10) GUT with $a_{24} \gg a_1, v_R$ (now) is
\begin{equation}
\Gamma^{SU(5)} : \Gamma^{\rm now}
\simeq 1: 136,
\end{equation}
for $A_L \simeq A_R$,
if the $X$ gauge boson masses are the same.
Since the naive prediction of $p\to \pi^0e^+$ partial lifetime in SU(5) GUT
is $\tau_p \sim 10^{36}$ years,
the prediction in the SO(10) with $a_{24} \gg a_1, v_R$
is $10^{34}$ years, which is on the current experimental bound at SK.


\section{SO(10) breaking vacua in the model}

We consider a superpotential for the GUT breaking Higgs fields ${\bf 45}_{\rm H}$ ($A$), ${\bf16}_{\rm H}$ ($\chi$) and $\overline{{\bf16}}_{\rm H}$ ($\bar{\chi}$),
\begin{equation}
W = \frac12 m_A A^2 + \frac{\lambda_1}{M_*} (A^2)^2 + \frac{\lambda_2}{M_*} A^4 + m_\chi \chi \bar \chi + \kappa A \chi \bar \chi,
\end{equation}
where we define $A^2 \equiv A_{ab} A_{ab}/2$, and $A^4 \equiv A_{ac} A_{ad} A_{bc} A_{bd}/2$
so that the multiplication of contraction of 2-anti-symmetric indices is removed by dividing by 2.
The superpotential in terms of canonically-normalized SM singlets
$a_1$, $a_{24}$ in $A$ and $v_R$ in $\chi$ (Clebsch-Gordan coefficients
for $\bf 16$ representation can be found in \cite{Nath:2001yj}) is given by
\begin{eqnarray}
W\vert_{\rm SM \ singlets}
&=& m_\chi v_R \bar v_R + \frac12 m_A (a_1^2 + a_{24}^2)
+ \sqrt5 \kappa \, a_1 v_R \bar v_R \nonumber\\
&&  + \frac{\lambda_1}{M_*} (a_1^2+a_{24}^2)^2 
+ \frac{\lambda_2}{5M_*} \left(a_1^4+ 6 a_1^2 a_{24}^2+ 2\sqrt{\frac23} a_1 a_{24}^3 + \frac76 a_{24}^4\right).
\end{eqnarray}
The $F$-flat conditions read
\begin{eqnarray}
&&a_1 = -\frac{m_\chi}{\sqrt5 \kappa},\label{a1}\\
&&a_{24} = \frac{m_\chi}{\sqrt5\kappa}a,\\
&&v_R \bar v_R = \frac{4m_\chi^3}{25\kappa^4 M_*} \left(\lambda_1(1+a^2)+\frac{\lambda_2}5(1+3a^2-\frac1{\sqrt6} a^3)\right)
+\frac{m_\chi m_A}{5\kappa^2},
\end{eqnarray}
where $a$ is a solution to
\begin{equation}
\lambda_1 + \frac35 \lambda_2 + \frac{5\kappa^2 m_A M_*}{4m_\chi^2}- \sqrt{\frac32}\frac{\lambda_2}5 a 
+ (\lambda_1 + \frac7{30} \lambda_2) a^2 =0.
\label{a24}
\end{equation}
In Eq.~(\ref{a1}), the condition $\partial W/\partial v_R = (m_\chi + \sqrt{5}\kappa a_1) \bar v_R$ fixes
the VEV of $a_1$ to be around $m_\chi$.
In Eq.~(\ref{a24}), on the other hand, the VEV of $a_{24}$ (proportional to $a$) is fixed by a balance 
between the quadratic mass term and the quartic non-renormalizable term, and
$|a|$ is large if $|m_A M_*/m_\chi^2| \gg 1$, in which case we obtain
\begin{eqnarray}
&&
a_{24}^2 \sim  -\frac{m_A M_*}{4(\lambda_1 + \frac7{30} \lambda_2)}, \qquad
v_R \bar v_R \sim \frac{\lambda_2}{5\sqrt{30} \kappa} \frac{m_A a_{24}}{\lambda_1 + \frac7{30} \lambda_2}.
\end{eqnarray}
Thus, vacua with $|v_R|,|a_1| \ll |a_{24}|$ are obtained\footnote{
If one adds a non-renormalizable term $(\chi \bar\chi)^2/M_*$ to the superpotential,
new vacua with $v_R \sim \sqrt{m_\chi M_*}$ appear.
However, the vacua we obtain in the main text remain stable with a correction of $O(v_R^2/M_*^2)$.
}
with a feasible assumption $m_\chi,m_A \ll M_*$.

\section{GUT-scale threshold corrections for the gauge coupling unification}

The gauge coupling unification conditions \cite{Hisano:1992jj} in SUSY SO(10) GUT are written as\footnote{
The SUSY particle threshold contributions in the respective equations are more precisely written as
 \begin{eqnarray}
-2 \ln \frac{m_{\rm SUSY1}}{m_Z}
&=& -\frac85 \ln \frac{\mu_H}{m_Z} - \frac25 \ln \frac{m_H}{m_Z}
+ 4 \ln \frac{M_{\tilde g}}{M_{\tilde w}} 
+ \frac35 \ln \frac{m_{\tilde q^c}^3 m_{\tilde d^c}^2 m_{\tilde e^c}}{m_{\tilde q}^4 m_{\tilde \ell}^2}, 
\\
8 \ln \frac{m_{\rm SUSY2}}{m_Z}
&= &
 4 \ln \frac{M_{\tilde g} M_{\tilde w}}{m_Z^2}
+ 3 \ln \frac{m_{\tilde q}^2}{m_{\tilde u} m_{\tilde e^c}},
\end{eqnarray}
where $\mu_H$, $m_H$ are higgsino and heavier Higgs masses,
$M_{\tilde g}$ and $M_{\tilde w}$ are gluino and wino masses.
From these equations,
one finds that
the colored Higgs mass is larger for a smaller ratio of $M_{\tilde g}/M_{\tilde w}$,
and the unification scale $M_G$ becomes smaller for heavier wino and gluino masses.
}

\begin{eqnarray}
&&-\frac{2}{\alpha_3(m_Z)} + \frac{3}{\alpha_2(m_Z)} - \frac{1}{\alpha_1(m_Z)}
= \frac1{2\pi} \left(\frac{12}5 \ln \frac{M_{H}}{m_Z} -2 \ln \frac{m_{\rm SUSY1}}{m_Z}
\right),
\\
&&-\frac{2}{\alpha_3(m_Z)} - \frac{3}{\alpha_2(m_Z)} + \frac{5}{\alpha_1(m_Z)}
= \frac1{2\pi} \left(36 \ln \frac{M_G}{m_Z} + 8 \ln \frac{m_{\rm SUSY2}}{m_Z}
\right),
\end{eqnarray}
\begin{eqnarray}
M_H &=& \frac{M_{Q}^4}{M_{E} M_{U}^3 } M_{H_C} \prod_i M_i^{l_A^i}, \\
M_G^6 &=& \frac{M_{X}^4 M_{E}^2 M_{U}^2}{M_{Q}^4} \prod_i M_i^{l_B^i},
\end{eqnarray}
where $M_{X,Q,U,E}$ are the SO(10) gauge boson masses which we have already defined,
 $M_{H_C}$ is the colored Higgs mass,
 and $i$ stands for the degree of physical modes under the SM decompositions. 
We define $l_A = \frac{5}{12} (2 l_3 - 3 l_2 +  l_1)$
and $l_B = \frac16 (2 l_3 + 3 l_2 - 5 l_1)$ where
$ l_i$ gives the beta coefficient contribution of the respective multiplet, $l_i=\Delta b_i^{\rm SUSY} $.
Because the would-be-Goldstone modes which are eaten by the gauge bosons
lack from the multiplets, 
we obtain 
 \begin{equation}
\sum_i l_A^i = 0, \qquad
\sum_i l_B^i = 2.
\end{equation}
The RGEs give
\begin{equation}
M_H \sim 10^{15}-10^{16} \ {\rm GeV}, \qquad
M_G \sim 2 \times 10^{16} \ {\rm GeV},
\end{equation}
and the GUT gauge boson and colored Higgs masses 
depend on threshold corrections of GUT-scale particles.

The $\bf45_{\rm H}$ and $\bf 16_{\rm H}$ representations are decomposed under SU(5) as
${\bf 45}_{\rm H} = {\bf 24} + {\bf 10} + \overline{\bf10} + {\bf 1}$.
and
${\bf 16}_{\rm H} = {\bf 10} + \overline{\bf 5} + {\bf1}$.
One linear combination of the ${\bf 10}$'s (and $\overline{\bf10}$'s) from
$\bf 45_{\rm H}$ and $\bf 16_{\rm H}$ ($\overline{\bf16}_{\rm H}$) is absorbed by the
gauge bosons $Q,U,E$.
For $|v_R| \ll |a_{24}|$, the linear combination to be absorbed mainly comes from $\bf45_{\rm H}$.
The other linear combination is a physical mode and we denote its components by $\chi_Q,\chi_U,\chi_E$
 (which respectively have the same SM charge as $Q,U,E$).
For $|m_A|,|m_\chi| \ll M_*$,
 their masses satisfy the ratio (see Appendix~A for the derivation)
\begin{equation}
M_{\chi_Q} : M_{\chi_U}: M_{\chi_E} \simeq 1: 4 : 6.
\label{physicalhiggsmasses}
\end{equation}
The $\bf24$ representation in $\bf45_{\rm H}$ contains a SU(3)$_c$ adjoint $({\bf8},{\bf1},0)$
and a SU(2)$_L$ adjoint $({\bf1},{\bf 3},0)$ as physical  modes.
Their masses can be calculated (using the minimization conditions) as
%
\begin{eqnarray}
M_8 &=& \frac23\frac{\lambda_2}{M_*} a_{24} (-3\sqrt6 a_1 + a_{24}),\\
M_3 &=& \frac23\frac{\lambda_2}{M_*} a_{24} (3\sqrt6 a_1 + 4a_{24}),\label{m3}
\end{eqnarray}
and when $|a_1| \ll |a_{24}|$, we find
\begin{eqnarray}
M_8 : M_3 \simeq 1:4.
\label{physicalhiggsmasses2}
\end{eqnarray}
The other physical Higgs modes are the ${\bf5}+\bar{\bf5}$
in ${\bf16}_{\rm H}+\overline{\bf16}_{\rm H}$.
We additionally introduce two $\bf10_{\rm H}$ representations to 
generate renormalizable Yukawa couplings that give fermion masses
after electroweak symmetry breaking.
Then, there are three heavy colored Higgs fields $H_C$, $H_{T_{1,2}}$, 
and two heavy Higgs doublets $H_{D_{1,2}}$ around the GUT scale.

Now the gauge coupling unification conditions are specified as
\begin{eqnarray}
M_H &=& \frac{M_{Q}^4}{M_{E} M_{U}^3 }  \frac{(M_{\chi_E} M_{\chi_U}^3)^{1/2}}{M_{\chi_Q}^2}
\left(\frac{M_8}{M_3}\right)^{5/2} \frac{M_{T_1} M_{T_2}}{M_{D_1} M_{D_2}} M_{H_C}, \label{mhrelation}\\
M_G^6 &=& \frac{ M_{E}^2 M_{U}^2}{M_{Q}^4} \frac{M_{\chi_Q}^2}{M_{\chi_E} M_{\chi_U}}M_{X}^4 M_8 M_3.\label{mgrelation}
\label{mg}
\end{eqnarray}
In the vacua with $|v_R|,|a_1|\ll |a_{24}|$,
we obtain from Eqs.~(\ref{gaugebosonmasses}),(\ref{physicalhiggsmasses}),(\ref{physicalhiggsmasses2}),
\begin{eqnarray}
&&\frac{M_{Q}^4}{M_{E} M_{U}^3 } \frac{(M_{\chi_E} M_{\chi_U}^3)^{1/2}}{M_{\chi_Q}^2}
\left(\frac{M_8}{M_3}\right)^{5/2}
\simeq 
\frac1{576},\\
&&
 \frac{M_{E}^2 M_{U}^2}{M_{Q}^4} \frac{M_{\chi_Q}^2}{M_{\chi_E} M_{\chi_U}} \simeq 24.
 \label{mg2}
\end{eqnarray}
Due to the factor $1/576$,
the colored Higgs mass can be much larger
than in the minimal SU(5) model.
As for the gauge boson mass,
in the minimal SU(5) model,
 one has $M_G^6=M_X^4 M_8 M_3$ and
 $M_8 = M_3 = \lambda M_X$
where $\lambda$ is proportional to the self-coupling of the SU(5) adjoint representation.
$\lambda$ is arbitrary unless it far exceeds $O(1)$,
and people often assume $\lambda\sim1$, which gives $M_X\sim M_G$.
In the current SO(10)-breaking vacua $|v_R|,|a_1|\ll |a_{24}|$,
 if we write $4M_8 \simeq M_3 = \rho M_X$,
 $\rho$ is always much smaller than 1
 because the masses of the SU(3)$_c$ adjoint and SU(2)$_L$ adjoint  
 particles are roughly $m_A$,
while the $X$ gauge boson mass is roughly $(m_A M_*)^{1/2}$.
To be specific, 
 we get from Eqs.~(\ref{mg}),(\ref{mg2}),
\begin{equation}
M_X \simeq M_G \frac1{(6\rho^2)^{1/6}},
\label{xgratio}
\end{equation}
 and from Eqs.~(\ref{mx}),(\ref{m3}),
\begin{equation}
\rho = \frac{M_3}{M_X} \simeq \frac{16}{5} \lambda_2 \frac{1}{g_5^2}\frac{M_X}{M_*}.
\end{equation}
Therefore we find
\bea
\rho\simeq2\left(\frac{\lambda_2}{g_5^2}\right)^{3/4}\left(\frac{M_G}{M_*}\right)^{3/4},
\eea
 which equals 0.1 for $\lambda_2=1$ and $M_*=2\times10^{18}$~GeV.
It follows that $M_X$ is a little larger than $M_G$.
Nevertheless, the $Q$ gauge boson satisfying $M_X:M_Q \simeq 5:1$
 enhances the dimension-6 proton decay width compared to the minimal SU(5) model.

%

To summarize, in the SO(10)-breaking vacua with $|a_{24}| \gg |a_1|, |v_R|$,
the colored Higgs is made heavier by the GUT-scale threshold corrections,
and the dimension-5 proton decay is suppressed compared to the minimal SU(5) model.
On the other hand,
the dimension-6 proton decay width is roughly 100 times enlarged
and we have $\tau_p \sim 10^{34}$~years, which is in the scope of HK.
\\

Suppression of the dimension-5 proton decay is also achieved by making the ratio of gluino and wino masses $M_{\tilde g}/M_{\tilde w}$ smaller,
 and enhancement of the dimension-6 proton decay is achieved by increasing their product $M_{\tilde g} M_{\tilde w}$,
 as seen from the SUSY particle threshold correction formulas.
Hence, in the high scale SUSY scenario, the dimension-6 proton decay 
is detectable at HK even in the minimal SUSY SU(5) model.
In contrast, in our SO(10)-breaking vacua $|a_{24}| \gg |a_1|, |v_R|$,
the GUT-scale threshold corrections enhance the dimension-6 proton decay width to a detectable level,
even if SUSY particle masses are a few TeV.
\\

We comment on the case when the ${\bf 16}_{\rm H}$ is replaced by 
$\bf 126_{\rm H}$ representation.
In this case, when a vacuum with $|a_1| \ll |a_{24}|$ is chosen,
$({\bf 6},{\bf 3},1/3)$ multiplet is about $1/3$ lighter than the other components in the representation.
Since this multiplet has $l_A=-33/2$, it gives a large threshold correction
and renders the colored Higgs too light.

\section{Numerical result}

In the previous section,
we have used 1-loop relations
to describe qualitative behaviors.
In this section,
we will show a numerical result using 2-loop RGE evolutions  \cite{Jones:1981we,Machacek:1983tz}.
In the result, we use the central value of the 5-flavor strong coupling,
$\alpha_s^{(5)} (M_Z)^{\overline{\rm MS}} = 0.1181 \pm 0.0011$ \cite{Tanabashi:2018oca}.
The colored Higgs mass is sensitive to the value of the strong coupling,
while the GUT gauge boson masses are less sensitive.
The proton lifetime is about $50\%$ larger if we use the value $+ 3\sigma$.
We assume
all the SUSY particle masses to be 2 TeV except for the wino mass, 
which is taken to be 500 GeV.

%


The decay width of $p \to \pi^0 e^+$ is \cite{Nath:2006ut}
\begin{equation}
\Gamma = \frac{\pi}{4} \frac{m_p \alpha_H^2}{f_\pi^2} (1+F+D)^2\alpha_U^2 \left[ A_{ R}^2 \left(\frac{1}{M_X^2}+\frac1{M_Q^2}\right)^2 + \frac{4A_{L}^2}{M_X^4} \right],
\end{equation}
where we use proton mass $m_p = 0.9383$ GeV, 
chiral Lagrangian parameters $F=0.46$, $D=0.80$, 
hadron matrix element for proton decay $\alpha_H = -0.014 \ {\rm GeV}^3$ at 2 GeV \cite{Aoki:2017puj},
decay constant $f_\pi = 0.1307$ GeV, 
renormalization factors down to 2 GeV, $A_L = 2.91$, $A_R = 2.78$ (The two-loop renormalization factors are calculated in \cite{Hisano:2013ege}).
From the SO(10) gauge coupling unification, we obtain $1/\alpha_U (M_X) = 4\pi/g_5^2 \simeq 25.2$.

Before presenting the main result, we show an estimate on the partial proton lifetime.
Under the approximations with
\begin{eqnarray}
M_X : M_3 : M_8 = 1: \rho : \rho/4,\\
M_{\chi_Q}:M_{\chi_U}:M_{\chi_E} = 1:4:6,
\end{eqnarray}
and
\begin{equation}
M_X : M_Q : M_U : M_E = 5: 1: 4:6,
\end{equation}
 the partial proton lifetime is found to be
\begin{equation}
\tau_p \simeq \left(\frac{0.1}{\rho}\right)^\frac{4}{3} \times 2.6 \times 10^{34} \ {\rm years}.
\label{taupestimate}
\end{equation}
As discussed in the previous section, $\rho \ll 1$ in the current SO(10)-breaking vacua because the VEV of 
 $a_{24}$ is roughly the geometrical average of $m_A$ and $M_*$
 while $M_3,M_8$ are roughly $m_A$, and we get $\rho\simeq0.1$ for $\lambda_2=1$ and $M_*=2\times10^{18}$~GeV.
 
It is interesting to compare the above estimate 
 with the prediction of the minimal SU(5) model.
In the minimal SU(5), we define $M_8 = M_3 = \lambda M_X$ where
$\lambda$ is proportional to the self-coupling of the adjoint field that breaks SU(5).
Then, the partial proton lifetime is found to be
\begin{equation}
\tau_p^{\rm SU(5)} \simeq \lambda{}^{-4/3} \times 5.5 \times 10^{35} \ {\rm years}.
\end{equation}
%
We observe that the partial lifetime decreases by $1/20$ in our SO(10)-breaking vacua compared to the minimal SU(5) model,
 for natural values of $\rho=0.1$ and $\lambda=1$.
 \\

The estimate for our SO(10)-breaking vacua, Eq.~(\ref{taupestimate}),
receives corrections 
from the small VEVs of $a_1,v_R$ that perturb the mass ratios.
In Table~1, we show precise numerical values.
Here, we fix 
$M_* = 2\times 10^{18}$ GeV, and take benchmark values for $\lambda_1, \lambda_2,\kappa$ and $m_\chi$.
%
We solve the $F$-flat conditions Eqs.~(\ref{a1})-(\ref{a24}) and
the unification conditions 
by varying $m_A$ and the colored Higgs mass.
Since Eq.~(\ref{a24}) is a quadratic equation,
there are two solutions.
If $m_A < 0$, $a_{24}$ is real
and the mass spectrum splits into two, both of which are tabulated.
If $m_A > 0$, $a_{24}$ is complex
and the two solutions yield the same mass spectrum in terms of the absolute values.
\begin{table}[H]
\center
\caption{The $p \to \pi^0 e^+$ partial lifetime
and the mass spectrum for various input values of
$\lambda_1,\lambda_2,\kappa$ and $m_\chi$.
The masses and VEV are in units of $10^{16}$ GeV.
From (i-A) to (ii-B), $m_A < 0$ and thus $a_{24}$ is real and the mass spectrum splits into two, for which
(i-A) and (ii-A) correspond to the cases with $a_{24}>0$,
and (i-B) and (ii-B) to the cases with $a_{24}<0$.
In (iii), $m_A> 0$ and thus $a_{24}$ is complex.
We change one of $\lambda_1,\lambda_2,\kappa$ in (iii$^\prime$),
(iii$^\prime{}^\prime$),
(iii$^\prime{}^\prime{}^\prime$).
}
\begin{tabular}{|c|c|c||c|c|c|c|c|c|c|}
\hline
& $(\lambda_1,\lambda_2,\kappa)$ & $m_{\chi}$ &$m_A$ & $M_X$ & $M_Q$ & $M_3$ & $M_8$ & $v_R$  &  $\tau_p (10^{34} \ {\rm years})$ \\
\hline
\hline
(i-A) & $(1,1,1)$ & 0.1 &$-0.47$ & $2.8$ & $0.55$ & $0.25$ & 0.068 & 0.24   & 3.4 \\
\hline
(i-B) & $(1,1,1)$ & 0.1 &$-0.46$ & $2.8$ & $0.61$ & $0.25$ & 0.057 & 0.25   & 4.7 \\
\hline
(ii-A) & $(1,1,1)$& 0.01 &$-0.48$ & $2.8$ & $0.59$ & $0.26$ & 0.069 & 0.25   & 4.3 \\
\hline
(ii-B) & $(1,1,1)$& 0.01 &$-0.44$ & $2.7$ & $0.57$ & $0.24$ & 0.059 & 0.24   & 3.7 \\
\hline
(iii) & $(1,1,1)$& 0.1&$0.46$ & $2.8$ & $0.58$ & $0.25$ & 0.063 & 0.24   & 4.0 \\
\hline
\hline
(iii$^\prime$) &$(1,0.5,1)$ &0.1& $0.58$ & $3.2$ & $0.67$ & $0.17$ & 0.043 & 0.22   & 7.2 \\
\hline
(iii$^\prime{}^\prime$) & $(0.1,1,1)$&0.1& $0.12$ & $2.8$ & $0.58$ & $0.25$ & 0.063 & 0.24   & 4.0 \\
\hline
(iii$^\prime{}^\prime{}^\prime$) & $(1,1,0.5)$&0.1& $0.48$ & $2.8$ & $0.62$ & $0.26$ & 0.066 & 0.36   & 5.1 \\
\hline
\end{tabular}
\end{table}
From (iii) and (iii$^\prime{}^\prime$) of Table~1, we find that
the mass spectrum is not sensitive to $\lambda_1$.
This is because 
the relation $|a_1|, |v_R| \ll |a_{24}|$
gives
$M_3/a_{24} \propto \lambda_2 a_{24}/{M_*}$.
Although $a_{24}$ depends on $\lambda_1$,
the ratio $M_3/M_X$ does not depend on $\lambda_1$
for $|a_1|, |v_R| \ll |a_{24}|$.
As a result, once $M_X$ is chosen to realize the gauge coupling unification,
the mass spectrum is almost independent of $\lambda_1$.
%
On the other hand, when $\lambda_2$ is smaller, 
the SU(3)$_c$ adjoint and SU(2)$_L$ adjoint particles
become lighter ($\rho = M_3/M_X$ is smaller),
and the proton lifetime becomes longer,
as seen from (iii) and (iii$^\prime$) of Table~1.
%
Consequently, the proton lifetime cannot be bounded from above theoretically.
%
Still, it is interesting that for $\lambda_2\sim1$,
the dimension-6 proton decay is detectable at HK.

In the benchmarks of Table~1,
the effective colored Higgs mass, $M_{H_T} = \frac{M_{T_1} M_{T_2} M_{H_C}}{M_{D_1} M_{D_2}} $,
is $2\times10^{17}$~GeV.
The relation $M_{H_T} > M_X$ is realized with a large coupling of $A H_1 H_2$
(see Appendix~A).
Since the dimension-5 proton decay amplitudes also 
depend on details of the Yukawa coupling unification,
we do not discuss the dimension-5 decay in this paper.

%
%



%


\section{Conclusion}

We have studied the dimension-6 proton decay in a SUSY SO(10) GUT with only
 ${\bf 45}_{\rm H}+{\bf 16}_{\rm H}+{\bf \overline{16}}_{\rm H}$ GUT-breaking Higgs fields.
Since the SU(5)-breaking VEV of ${\bf 45}_{\rm H}$ is induced by the Planck-suppressed, quartic superpotential for ${\bf 45}_{\rm H}$,
 this VEV is larger than the SU(5)-conserving VEVs.
This results in a $1/5$ suppression of the $Q({\bf 3},{\bf2},1/6)$ gauge boson mass compared to the $X({\bf 3},{\bf2},-5/6)$ gauge boson mass.
On the other hand, the masses of the SU(3)$_c$ adjoint and SU(2)$_L$ adjoint particles 
from {\bf 45}$_{\rm H}$ are much smaller than the $X$ gauge boson mass
and this enhances the latter when the unification condition is fulfilled.
Still, the mass of the $Q$ gauge boson can be below $0.6\times10^{16}$~GeV (for $\lambda_2=1$ and $M_*=2\times10^{18}$~GeV)
 and can thus give rise to a fast dimension-6 proton decay detectable at Hyper-Kamiokande.
\\

\appendix

\section{Mass spectrum}

${\bf 45}_{\rm H}(A)$ and ${\bf 16}_{\rm H}(\chi)$ representations are decomposed under SU(5) as
${\bf 45}_{\rm H} = {\bf 24} + {\bf 10} + \overline{\bf10} + {\bf 1}$
and
${\bf 16}_{\rm H} = {\bf 10} + \overline{\bf 5} + {\bf1}$.
One linear combination of the ${\bf 10}$'s (and $\overline{\bf10}$'s) from
$\bf 45_{\rm H}$ and $\bf 16_{\rm H}$ ($\overline{\bf16}_{\rm H}$) is absorbed by GUT
gauge bosons, $Q :({\bf3},{\bf2},1/6)$, $U: (\bar{\bf3},{\bf1},-2/3)$, and $E: ({\bf 1},{\bf 1},1)$.
The other linear combination yields physical modes 
$\chi_Q :({\bf3},{\bf2},1/6)$, $\chi_U: (\bar{\bf3},{\bf1},-2/3)$, and $\chi_E: ({\bf 1},{\bf 1},1)$.
The mass matrix of each component of the ${\bf10}+\overline{\bf10}$'s can be written as
\begin{equation}
\left( \begin{array}{cc}
M_{AA} & M_{A\bar\chi} \\
M_{\chi A} & M_{\chi\bar\chi}
\end{array}
\right),
\end{equation}
where
\begin{eqnarray}
M_{AA} &=& m_A + 4a_1^2 \frac{\lambda_1+\frac15\lambda_2}{M_*} + 4a_{24}^2 \frac{\lambda_1}{M_*}+
\frac45 \frac{\lambda_2}{M_*} \left(\frac{C_{Q,U,E}}{\sqrt6} a_1 a_{24} + \frac{D_{Q,U,E}}{6} a_{24}^2\right), \\
M_{A\bar\chi} & = &
2 \kappa v_R, \\
M_{\chi A} & = &
2 \kappa \bar v_R, \\
M_{\chi\bar\chi} &=&
m_\chi + \frac1{\sqrt5} \kappa a_1 - \sqrt{\frac{2}{15}} \kappa \, C_{Q,U,E} \, a_{24} 
\end{eqnarray}
where $(C_Q,C_U,C_E) = (1,-4,6)$ and $(D_Q,D_U,D_E) = (19,4,9)$.
We can verify that one eigenvalue is zero when the $F$-flat conditions are used.
The mass of the physical mode is $M_{AA} + M_{\chi\bar\chi}$.
In the limit with $m_A, m_\chi \ll M_*$, $M_{\chi\bar\chi}$ dominates,
 but $M_{AA}$ can be non-negligible for $\chi_Q$ due to the large factor $D_Q/C_Q$.
Using the minimization condition,
we obtain
$M_{AA} \simeq m_A(7-D_{Q,U,E})\lambda_2/( 30\lambda_1 + 7\lambda_2)$
for $m_A, m_\chi \ll M_*$.


The masses of isospin doublet and color triplet Higgses are obtained from the superpotential
\begin{equation}
W_H = M_{ij} H_i H_j + \lambda_H A H_1 H_2 + \lambda_\chi^i H_i \chi \chi + \bar\lambda_\chi^i H_i \bar\chi \bar\chi ,
\end{equation}
and the mass term is
\begin{equation}
(
\begin{array}{ccc}
 H_1^{\bf5} & H_2^{\bf5}
  & \bar\chi^{\bf5}
\end{array}
) M_{H_{D,T}} \left(
 \begin{array}{c}
  H_1^{\bar {\bf5}} \\ H_2^{\bar {\bf5}} \\ \chi^{\bar {\bf5}}
 \end{array}
\right),
\end{equation}
\begin{equation}
M_{H_{D,T}} = \left(
\begin{array}{ccc}
M_{11} & M_{12} +  {\lambda_H} A_{D,T} & \lambda_\chi^1 v_R\\
M_{12} -  {\lambda_H} A_{D,T} & M_{22} & \lambda_\chi^2 v_R\\
\bar\lambda_\chi^1 \bar v_R & \bar\lambda_\chi^2 \bar v_R & m_\chi + \frac{\kappa}{\sqrt5} (-3 a_1 + \sqrt{\frac2{3}} c_{D,T} a_{24})\\
\end{array}
\right),
\end{equation}
where
\begin{equation}
A_{D,T} = \frac{1}{\sqrt5} (a_1 + \frac{1}{\sqrt6} c_{D,T} a_{24})
\end{equation}
and
$(c_D,c_T)= (3,-2)$.
The doublet-triplet splitting needs fine-tuning.
Without loss of generality, $\lambda_\chi^1$ is set to zero by a rotation of $(H_1,H_2)$.
In this basis, by the fine-tuning $M_{11} = M_{12} + \lambda_H A_D= 0$, we have one pair of doublets massless.
$H_1$ in this basis should dominantly give the large top quark Yukawa coupling.
The mass of the corresponding triplet is roughly
$\sim 5/3 \lambda_H A_D$ 
for $|a_1|, |v_R|\ll |a_{24}|$.
\\

\section{Renormalizable model obtained by employing ${\bf 54}_{\rm H}$}

In the main text, we have considered the model with ${\bf45}_{\rm H} + {\bf 16}_{\rm H} + \overline{\bf16}_{\rm H}$
and with non-renormalizable quartic terms of $\bf 45_{\rm H}$. 
In this appendix, for readers who prefer renormalizable models,
we show that a renormalizable superpotential with $\bf54_{\rm H}$ 
(whose SM singlet component is denoted by $E$)
can also provide the wanted vacua where $|a_{24}| \gg |a_1|, |v_R|$ (and $|a_{24}| \gg|E|$).

The superpotential for the SM singlets is
\begin{eqnarray}
W\vert_{\rm SM \ singlets} 
&=& m_\chi v_R \bar v_R + \frac12 m_A (a_1^2 + a_{24}^2)
+\frac12 m_E E^2 +  \sqrt5\kappa \, a_1 v_R \bar v_R \\
&& +\frac{\kappa_1}3 E^3 + 
\kappa_2 E \left(\sqrt{6} a_1 a_{24} + \frac12 a_{24}^2 \right)
\end{eqnarray}
From the $F$-flat conditions, we obtain
\begin{equation}
a_1 = -\frac{m_\chi}{ \sqrt5\kappa}, \quad
a_{24} = \sqrt{\frac65}\frac{m_\chi}{\kappa}  a,
\quad
v_R\bar v_R = \frac{m_\chi m_A}{5 \kappa^2} \frac{6a^2+a-1}{a-1},
\quad
E =  -\frac{m_A}{\kappa_2} \frac{a}{a-1},
\end{equation}
where 
%
$a$ is a solution of the following equation:
\begin{equation}
m_E =  \kappa_1 \frac{m_A}{\sqrt5\kappa} \frac{a}{a-1} +  \frac{3 \kappa_2^2m_\chi^2}{5\kappa^2 m_A} (a-1)(a-2).
\end{equation}
Vacua with $|E|,|a_1|, |v_R| \ll |a_{24}|$ are obtained by assuming
$m_\chi, m_A \ll m_E$, which gives
\begin{equation}
a_{24} \simeq  \frac{\sqrt{2m_A m_E}}{\kappa_2}.
\end{equation}

The $\bf54_{\rm H}$ is decomposed as ${\bf 54}_{\rm H} = {\bf 24} + {\bf15} + \overline{\bf15}$
under SU(5).
%
The mass matrices of the adjoint representations after SU(5) breaking
are
\begin{equation}
\left(\begin{array}{cc}
 m_A - 2 \kappa_2 E & -2 \kappa_2 a_{24} + \sqrt6 \kappa_2 a_1  \\
 -2 \kappa_2 a_{24}+ \sqrt6 \kappa_2 a_1  & m_E - 4 \kappa_1 E
\end{array}\right)
\end{equation}
for $({\bf8},{\bf1},0)$
and
\begin{equation}
\left(\begin{array}{cc}
 m_A +3 \kappa_2 E & 3 \kappa_2 a_{24} + \sqrt6 \kappa_2 a_1  \\
 3 \kappa_2 a_{24}+ \sqrt6 \kappa_2 a_1  & m_E +6 \kappa_1 E
\end{array}\right)
\end{equation}
for $({\bf1},{\bf3},0)$.
For $a\gg 1$, $\kappa_2 E \simeq  -m_A$,
and we obtain the masses of the lighter adjoint fields (using $2 m_Am_E \simeq (\kappa_2 a_{24})^2$)
as
$M_8 \simeq (3-4\times2)m_A =-5 m_A$
and $M_3 \simeq (-2-9\times2)m_A = -20 m_A$, and hence
$M_8 : M_3 \simeq 1:4$.
We have thus verified that the mass ratio $M_8: M_3$ is the same as the model with the non-renormalizable 
terms, which is obtained by integrating out $\bf54_{\rm H}$.

\section*{Acknowledgement}
This work is partially supported by Scientific Grants by the Ministry
of Education, Culture, Sports, Science and Technology of Japan, Nos.~17K05415, 18H04590 and 19H051061 (NH), and No.~19K147101 (TY).


\end{document}